\documentclass[a4paper,10pt]{article} %% don't forget char=12
\input epsf.tex
\usepackage{graphics}

\newcommand{\be}{\begin{equation}}
\newcommand{\ee}{\end{equation}}
\newcommand{\ba}{\begin{eqnarray}}
\newcommand{\ea}{\end{eqnarray}}
\newcommand{\ban}{\begin{eqnarray*}}
\newcommand{\ean}{\end{eqnarray*}}

\newcommand{\ket}[1]{\mbox{$ | #1 \rangle $}}
\newcommand{\bra}[1]{\mbox{$ \langle #1 | $}}

\newcommand{\si}{\sigma}

\newcommand{\one}{\leavevmode\hbox{\small1\normalsize\kern-.33em1}}

\begin{document}

\title{Partial list of bipartite Bell inequalities with four binary settings}
\author{Nicolas Brunner, Nicolas Gisin \\  Group of Applied Physics, University of
Geneva \\ nicolas.brunner@physics.unige.ch}
\date{\today}
\maketitle

\abstract{We give a partial list of 26 tight Bell inequalities for
the case where Alice and Bob choose among four two-outcome
measurements. All tight Bell inequalities with less settings are
reviewed as well. For each inequality we compute numerically the
maximal quantum violation, the resistance to noise and the minimal
detection efficiency required for closing the detection loophole.
Surprisingly, most of these inequalities are outperformed by the
CHSH inequality.}

\vspace{20pt}

Finding all the Bell inequalities for a given number of
measurement settings and outcomes is a difficult problem
\cite{pito}. Even for the case of binary settings (two-outcome
settings), few is known. All tight Bell inequalities, i.e. facets
of the local polytope \cite{pito,tsirelson2}, have been listed for
the following cases: 2222, M222 ($M\geq3$), 3322 and 4322, where
$ijmn$ refers to the situation where Alice chooses among $i$
settings with $m$ outcomes and Bob among $j$ settings with $n$
outcomes. For the cases 2222 and $M222$ ($M\geq 3$), there is only
one Bell inequality \cite{fine,dan}: the famous
Clauser-Horne-Shimony-Holt (CHSH) \cite{chsh} inequality

\ba \label{chsh} \textrm{CHSH} \equiv
\begin{array}{r|rr} & -1&0\\\hline  -1&1&1\\ 0&1&-1\\
\end{array} \leq 0 \quad.  \ea Here the notation represents the coefficients that are put in
front of the probabilities, according to \ba
\begin{array}{c|c} & P(r_B=0|y)\\\hline P(r_A=0|x)&
P(r_A=r_B=0|xy)
\end{array} \quad , \ea where $x$ ($y$) denotes the measurement setting of Alice (Bob) and
$r_A$ ($r_B$) its result.

In the 3322 case, only one new inequality \cite{dan} appears

\ba I_{3322} &\equiv& \begin{array}{r|rrr} & -1&0&0\\\hline -2&1&1&1\\ -1&1&1&-1\\ 0&1&-1&0\\
\end{array}\leq 0
\ea Note that by first inverting the output of Alice's first
setting as well as Bob's second and third settings, and then
relabelling the settings, one gets a symmetric version of
$I_{3322}$
\ba \tilde{I}_{3322} &\equiv& \begin{array}{r|rrr} & -1&-1&0\\\hline -1&0&1&1\\ -1&1&-1&1\\ 0&1&1&-1\\
\end{array}\leq 0 \quad .
\ea

For the case 4322, there are three new Bell inequalities
\cite{dan}:

\ba \nonumber  I_{4322}^1 \equiv
\begin{array}{r|rrr} & -1&0&0\\\hline  -2&1&1&1 \\ -1&1&-1&1\\
-1&1&1&-1\\0&1&-1&-1\\\end{array}\leq 0 \quad I_{4322}^2 \equiv
\begin{array}{r|rrr} & -2&-1&0\\\hline  -1&1&1&1\\ 0&0&1&-1\\
0&1&-1&0\\0&1&0&-1\\\end{array}\leq 0  \quad   I_{4322}^3 \equiv
\begin{array}{r|rrr} & -1&-1&0\\\hline  -2&2&1&1 \\ -1&-1&1&1\\
0&0&1&-1\\0&1&-1&-1\\\end{array}\leq 0 \ea

For more settings very little is known.

In this note we give a partial list of 26 inequivalent tight Bell
inequalities for the 4422 case. We believe that this list is not
exhaustive, in the sense that it does probably not contain all the
inequivalent facets of the 4422 local polytope. Note that there
exist algorithms that can list all Bell inequalities for a given
configuration (number of settings and outcomes) \footnote{Such
algorithms can be found on the internet: see for example
www.math.tu-berlin.de/polymake,
www.zib.de/Optimization/Software/Porta,
www.cs.mcgill.ca/˜fukuda/soft/cdd home/cdd.html.}. However, these
methods can still not deal with the complexity of the 4422 case.
To find the inequalities presented here, we have performed a
numerical research, inspired from the form of the inequalities
mentioned above. More precisely, we constructed all tables of the
form

\ba I &\equiv& \begin{array}{r|rrrr} & M(B_0)&M(B_1)&M(B_2)&M(B_3)\\\hline M(A_0)&C(A_0,B_0)&C(A_0,B_1)&C(A_0,B_2)&C(A_0,B_3)\\
M(A_1)&C(A_1,B_0)&C(A_1,B_1)&C(A_1,B_2)&C(A_1,B_3)\\ M(A_2)&C(A_2,B_0)&C(A_2,B_1)&C(A_2,B_2)&C(A_2,B_3)\\
M(A_3)&C(A_3,B_0)&C(A_3,B_1)&C(A_3,B_2)&C(A_3,B_3)\\\end{array}
\quad , \ea where all correlation coefficients $C(A_i,B_j)$ are
integers between $-2$ and $+2$, and all marginal coefficients
$M(A_i)$, $M(B_j)$ are integers chosen such that $-3 \leq M(A_0)<
M(A_1) \leq M(A_2)\leq M(A_3)=0$, and a similar relation for
coefficients $M(B_j)$. For each table we derive the local bound
$L$, by finding the maximum over all local deterministic
strategies --- there are $m^i n^j=256$ deterministic strategies in
the 4422 case. Finally we check if the obtained inequality ($I\leq
L$) is a facet of the local polytope or not. This is done by
computing the dimension of the subspace spanned by all
deterministic strategies saturating the inequality; if this
subspace is found to be an hyperplane (i.e. its dimension is
$d-1$, $d$ being the dimension of the no-signaling polytope
\cite{barrett}) then the inequality is tight.

As mentioned above, we find 26 inequivalent tight Bell
inequalities, which can be found at the end of this paper (see
Appendix A). Eight of these inequalities were already known, the
others are new to our knowledge. More specifically, $I_{4422}^1$
was presented in \cite{dan} , $I_{4422}^2$ in \cite{danPrivate} ,
$A_{5,6}$ and $AII_{1,2}$ in \cite{avis06} , $AS_{1,2}$ in
\cite{gisinAS} and independently in \cite{AS_WW}. Note that there
are only two correlation inequalities in this list, $AS_{1,2}$,
which Avis and co-workers have shown to be the only ones for the
4422 case \cite{avis44corr}.

Most of these 4422 inequalities can be written in a symmetric
form, as the CHSH and $I_{3322}$ inequalities. Among the 26
inequalities, only six cannot be written in such a form:
$I_{4422}^2$, $AII_{2}$, $I_{4422}^3$, $I_{4422}^5$, $I_{4422}^6$
and $I_{4422}^7$. It is not clear wether this symmetry is
meaningful or not.

Now we characterize all these inequalities. In particular we find
numerically the maximal quantum violation for a pure entangled
state of two qubits of the form \ba\label{NME}
\ket{\psi(\theta)}=\cos{\theta}\ket{00}+\sin{\theta}\ket{11} \quad
.\ea The maximal violation is found for the state
$\ket{\psi(\theta_{max})}$. We also give the resistance to noise
and the minimal detection efficiency required to close the
detection loophole. Results are summed up in table 1. Next we give
some comments on these results.

\begin{center}
\begin{table}[t!] \label{mainRES}
\begin{tabular}{||c||c|c|c|c|c||}
 \hline\hline

 Bell Inequality    & Quantum Max. Violation & $\theta_{max}/\pi$ & $w_{max}$  & $w$ & $\eta$ (Detection Loophole) \\ \hline\hline
 CHSH        &  $\frac{1}{\sqrt{2}}-\frac{1}{2} \approx 0.2071 $       &      $0.2500$       &  0.7071  &      $\frac{1}{\sqrt{2}}\approx0.7071$      &   $\frac{2}{\sqrt{2}+1}\approx0.8284$    \\
 $I_{3322}$  &      0.2500      &       0.2500         &    0.8000        &   0.8000  & 0.8284 \\
 $I_{4322}^1$   &    0.2361         &       0.2668        &    0.8640 &  0.8660           & 0.8761\\
 $I_{4322}^2$   &     0.2596        &     0.2749           &   0.8280    &   0.8333     &   0.8685   \\
 $I_{4322}^3$   &     0.4365        &        0.2500        &       0.7746     &  0.7746  & 0.8514 \\
 \hline  \hline
 $I_{4422}^1$   &       0.1970 &     0.2644     &    0.8988  &     0.9000     &    0.8571  \\
 $I_{4422}^2$   &       0.6214      &    0.2479      &    0.7630    & 0.7630  &    0.8443  \\
 $A_{5}$   &      0.4353       &    0.2450  &    0.7751       &  0.7752 &  0.8214 \\
 $A_{6}$   &      0.2321       &        0.2500        &    0.8829       &   0.8829 &  0.8373 \\
 $AS_{1}$   &     0.5412        &    0.2500     &   0.7348     &     0.7348    &   0.8472   \\
 $AS_{2}$   &     0.8785        &    0.2500     &  0.7400 &      0.7400  &    0.8506  \\
 $AII_{1}$   &    0.6055         &      0.2564         &   0.7676      & 0.7679    &  0.8323\\
 $AII_{2}$   &    0.5000         &   0.2500    &   0.8000       &  0.8000   &  0.8508\\
 $I_{4422}^3$   &       0.2380      &       0.2257     &     0.8630     &  0.8660 &  0.8761 \\
 $I_{4422}^4$   &     $\frac{1}{\sqrt{2}}-\frac{1}{2} \approx 0.2071$  & 0.2500  &  0.7071  &  $\frac{1}{\sqrt{2}}\approx0.7071$    &   $\frac{2}{\sqrt{2}+1}\approx0.8284$  \\
 $I_{4422}^5$   &     0.4365        &     $0.2500$           &     0.7746     &  0.7746  & 0.8514 \\
 $I_{4422}^6$   &        0.4495     &     $0.2500$           &     0.8165     &  0.8165  &  0.8697\\
 $I_{4422}^7$   &       1.4548      &      0.2622          &       0.7937    &  0.7949  & 0.8405 \\
 $I_{4422}^8$   &       0.4206   & 0.2457   & 0.8560   & 0.8561   & 0.8858  \\
 $I_{4422}^9$   &        0.4617  &  0.2648  &  0.8441  &  0.8455  &  0.8392  \\
 $I_{4422}^{10}$   &     0.6139  &  0.2538  &  0.8175  &  0.8176  &  0.8458  \\
 $I_{4422}^{11}$   &     0.6384  &  0.2444  &  0.7790  &  0.7792  &  0.8474  \\
 $I_{4422}^{12}$   &     0.6188  &  0.2404  &  0.7843  &  0.7849  &  0.8382  \\
 $I_{4422}^{13}$   &     0.2500  &  0.2500  &  0.8889 &  0.8889  &  0.8944  \\
 $I_{4422}^{14}$   &     0.4103  & 0.3790   & 0.8298   & 0.8310   & 0.8523 \\
 $I_{4422}^{15}$   &     0.2500  & 0.2500   &0.8889   & 0.8889   & 0.8944 \\
 $I_{4422}^{16}$   &     0.2407  &  0.2810  &  0.8791  &  0.8829  &  0.9009 \\
 $I_{4422}^{17}$   &     0.6714  &  0.2503  &  0.7883  &  0.7883  &  0.8611\\
$I_{4422}^{18}$   &    0.1812 &0.2498 & 0.9575& 0.9623 & 0.9575  \\
$I_{4422}^{19}$   &    0.4307 &0.2500 & 0.8745 & 0.8745 &0.8870 \\
$I_{4422}^{20}$   &  0.3056   & 0.3036 & 0.9075 &0.9231 & 0.8990\\
\hline\hline
\end{tabular}
\caption{All tight two-outcome Bell inequalities known to date,
with a number of settings up to four on each side. The list is
proven to be complete for inequalities with up to four settings
for Alice and three settings for Bob. Here we give a partial list
of 26 Bell inequalities for the $4422$ case. For each inequality
we find numerically the quantum state that achieves the largest
violation $\ket{\psi(\theta_{max})}$, and we give its resistance
to noise $w_{max}$. For the maximally entangled state, we provide
the resistance to noise $w$, as well as the detection efficiency
$\eta$ required to close the symmetric detection loophole. All
quantities are computed for two-qubit systems and non-degenerate
measurements (except for $I_{4422}^{4}$, see text).}
\end{table}
\end{center}

\textit{Quantum violation.} To find the largest quantum violation
we optimize over projective non-degenerate von Neumann
measurements. So each measurement setting for Alice (Bob) is
characterized by a vector $\vec{a}_x$ ($\vec{b}_y$) on the Bloch
sphere. Indeed one has $p(00|xy)=Tr(A_x
 \otimes B_y \ket{\psi(\theta)} \bra{\psi(\theta)})$ where $A_x \equiv \frac{\one +
\vec{a}_x\vec{\si}}{2}$ and a similar expression for $B_y$.

The CHSH and $I_{3322}$ inequalities are maximally violated by the
maximally entangled state, i.e. $\theta_{max}=\frac{\pi}{4}$.
Curiously for more settings, the maximal violation is often
obtained for a non-maximally entangled state (see Table 1). This
was already known to be the case for Bell inequalities with more
than two outcomes \cite{toni}, but is new, at least to our
knowledge, in the case of binary outcomes.

The inequality $I_{4422}^4$ has an astonishing feature. It
requires degenerate von Neunmann measurements (i.e. measuring the
identity) for two settings on each side, in order to be violated
by the maximally entangled state. In other words the singlet
cannot violate this inequality if one considers only
one-dimensional projectors. It is easily seen that by forgetting
(i.e. measuring the identity) Alice's second and third
measurements and Bob's first and fourth measurement one gets the
CHSH inequality (\ref{chsh}) which is indeed violated by the
maximally entangled state.

\textit{Resistance to noise.} Consider the Werner state $\rho_w =
w \ket{\phi^+}\bra{\phi^+} + (1-w) \frac{\one}{4}$. The resistance
to noise is defined as the amount of noise $(1-w)$ that can be
added to the maximally entangled state, such that the global state
ceases to violate the inequality. In case the maximal violation is
obtained for a partially entangled state
($\theta_{max}\neq\frac{\pi}{4}$), we also compute the resistance
to noise for this state ($\ket{\psi(\theta_{max})}$). We note it
$w_{max}$. Surprisingly the simplest of the inequalities presented
here, the CHSH inequality, is still the most robust against noise.

\textit{Symmetric detection loophole.} We compute the threshold
detection efficiency required to close the detection loophole in
the symmetric case, where Alice and Bob have the same detection
efficiency ($\eta_A = \eta_B \equiv \eta$). The threshold
efficiency $\eta$ is computed for the maximally entangled state.
Since we restrict our study to binary outcomes, Alice and Bob must
define a strategy for the case of non detection; for example they
can choose to output always "0" whenever they get no detection.
Here we optimize over the strategy of non detection (see
\cite{AsymDetLoop} for more details). We find that inequality
$A_5$ of Ref. \cite{avis06} allows a slightly lower detection
efficiency compared to CHSH; $\eta=82.14\%$ for $A_5$ versus
$\eta=82.84\%$ for CHSH. This is the best value known to date for
two-outcome Bell inequalities. Note that Ref. \cite{zoology}
presents a three-outcome inequality achieving a similar efficiency
($\eta=82.14\%$) \footnote{Though these two inequalities lead to
very similar thresholds, they are clearly different, since $A_5$
is a two outcome inequality while the inequality of Ref.
\cite{zoology} is a three-outcome inequality.}. The efficiency can
even be slightly lowered ($\eta=81.65\%$) by considering
inequalities sensitive to the source's efficiency \cite{zoology}.
Let us also remind that considering partially entangled states
helps to lower the threshold efficiency for CHSH, as shown by
Eberhard \cite{eberhard}. For inequality $A_5$ the threshold
efficiency decreases as long as the state is close to maximally
entangled but starts to increase again for very partially
entangled states. Thus $A_5$ allows to lower the threshold
efficiency compared to CHSH, only for states close to maximally
entangled ($\frac{\pi}{4.48} < \theta\leq \frac{\pi}{4}$).

\textit{Asymmetric detection loophole.} The study of the detection
loophole in asymmetric Bell tests
\cite{CabelloDetLoop,AsymDetLoop} is interesting, especially
considering atom-photon entanglement \cite{blinov}. The interest
of such systems come from the very high efficiency of the atomic
measurement (close to one), which allows to detect the photon with
a much smaller efficiency, compared to the symmetric case. It is
shown in Ref. \cite{AsymDetLoop} that, if the atom is always
detected, an efficiency as low as $43\%$ can be tolerated for the
photon detection, using the inequality $I_{3322}$ and very
partially entangled states. Two inequalities mentioned in the
present paper ($I_{4322}^2$ and $I_{4422}^3$) slightly improve
this result. In particular $I_{4422}^3$ can tolerate an efficiency
of $\sim 42,9\%$ in the limit of a very partially entangled state.

\textit{More settings.} Finally we mention some known Bell
inequalities with more than four two-outcome settings on each
side. In Ref. \cite{dan} a family of inequalities $I_{NN22}$ has
been derived for any number of settings $N$. These inequalities
generalize the inequalities $I_{3322}$ and $I_{4422}^1$ mentioned
in this paper. In Ref. \cite{avis06} two hundred millions of Bell
inequalities, with up to eight settings on each side, have been
found. See also \cite{gisinAS} for correlation inequalities with
more than four settings.

\textit{Conclusion.} In this paper we have presented 26 tight Bell
inequalities for the 4422 case. This list is, most probably, not
exhaustive. For each of these inequalities, we have computed the
largest quantum violation, the resistance to noise and the minimal
detection efficiency required to close the detection loophole in a
symmetric configuration. All these quantities have been computed
for two-qubit systems.

Finally, let us point out the astonishing power of the CHSH
inequality, the simplest Bell inequality. It clearly outperforms
most of the more complex inequalities presented here. To date it
is still the most robust against noise for qubit Werner states. It
is also the best for closing the detection loophole, except for
states close to maximally entangled, where inequality $A_5$ is
slightly better, as shown above. The largest improvement on CHSH
is found for the asymmetric detection loophole, where inequalities
$I_{3322}$ and $I_{4322}^2$ can tolerate much lower detection
efficiencies for the photon. This apparent superiority of the CHSH
inequality is surprising, since one may expect that considering
more data, i.e. having more settings to test, would help against
noise as well as for closing the detection loophole. Indeed we
have analyzed here only a partial list of four settings Bell
inequalities and a very efficient inequality might still be found.

The authors thank M.M. Wolf, K.F. P\'al, and T. V\'ertesi for
useful comments, and acknowledge financial support from the Swiss
NCCR "Quantum Photonics" project and the EU project QAP (N0.
IST-FET FP6-015848).

\textit{Note added in proof.} Recently P\'al and V\'ertesi
\cite{palvertesi} studied the violation of the Bell inequalities
presented here for higher dimensional systems as well as for
degenerate measurements.

\bibliographystyle{prsty}
\bibliography{H:/BIB/thesis}
%\bibliography{thesis}

%We also give the resistance to noise $p_{noise}$ for the state
%$\ket{\psi(\theta_{max})}$, as well as the resistance to noise for
%the maximally entangled state, $p_{noise}^{ME} $. Finally we give
%the minimal detection efficiency $\eta\equiv\eta_A=\eta_B$ that
%allows to close the detection loophole for the state
%$\ket{\psi(\theta_{max})}$.

\section{Appendix A: partial list of 4422 Bell inequalities}

\ba \nonumber  I_{4422}^1 \equiv
\begin{array}{r|rrrr} & -1&-1&-1&0\\\hline  -1&0&0&1&1 \\ -1&0&1&-1&1\\
-1&1&-1&-1&1\\0&1&1&1&-1\\\end{array}\leq 0 \quad \quad I_{4422}^2
\equiv
\begin{array}{r|rrrr} & -3&-1&0&0\\\hline  -2&2&1&2&0\\ -1&1&1&-1&1\\
0&2&-2&-1&0\\0&1&1&-1&-1\\\end{array}\leq 0  \\\nonumber A_{5}
\equiv
\begin{array}{r|rrrr} & -1&-1&-1&0\\\hline  -1&0&1&1&1 \\ -1&1&1&1&-1\\
-1&1&1&-1&0\\ 0&1&-1&0&0\\\end{array}\leq 0 \quad \quad A_{6}
\equiv
\begin{array}{r|rrrr} & -1&-1&0&0\\\hline  -1&1&1&0&1\\ -1&1&0&1&-1\\
0&0&1&-1&-1\\0&1&-1&-1&-1\\\end{array}\leq 0 \\\nonumber AS_{1}
\equiv
\begin{array}{r|rrrr} & -2&-1&0&0\\\hline  -2&1&1&1&1 \\ -1&1&1&1&-1\\
0&1&1&-2&0\\0&1&-1&0&0\\\end{array}\leq 0 \quad \quad AS_{2}
\equiv
\begin{array}{r|rrrr} & -3&-1&-1&0\\\hline  -3&1&1&2&2\\ -1&1&2&1&-2\\
-1&2&1&-2&1\\0&2&-2&1&-1\\\end{array}\leq 0   \\\nonumber AII_{1}
\equiv
\begin{array}{r|rrrr} & -1&-1&-1&0\\\hline  -1&-1&1&1&1 \\ -1&1&0&2&-1\\
-1&1&2&-1&-1\\0&1&-1&-1&0\\\end{array}\leq 0 \quad \quad AII_{2}
\equiv
\begin{array}{r|rrrr} & -3&-1&0&-1\\\hline  -1&2&1&1&-1\\ -1&1&2&-1&1\\
0&1&-1&-1&1\\0&1&-1&0&0\\\end{array}\leq 0     \\\nonumber
I_{4422}^3 \equiv
\begin{array}{r|rrrr} & -2&-1&-1&0\\\hline  -1&1&1&1&1 \\ 0&0&1&0&-1\\
0&1&-1&1&-1\\0&1&0&-1&0\\\end{array}\leq 0 \quad \quad I_{4422}^4
\equiv
\begin{array}{r|rrrr} & -1&-1&0&0\\\hline  -1&1&1&1&-1\\ -1&1&1&-1&1\\
0&1&-1&-1&-1\\0&-1&1&-1&-1\\\end{array}\leq 0    \\\nonumber
I_{4422}^5 \equiv
\begin{array}{r|rrrr} & -2&-1&0&0\\\hline  -1&1&0&1&0 \\ -1&1&1&-1&1\\
0&1&-1&0&0\\0&1&1&-1&-1\\\end{array}\leq 0 \quad \quad I_{4422}^6
\equiv
\begin{array}{r|rrrr} & -2&-1&-1&0\\\hline  -1&1&-1&1&1\\ -1&1&1&-1&1\\
0&1&-1&1&-1\\0&1&1&-1&-1\\\end{array}\leq  0 \\\nonumber
I_{4422}^7 \equiv
\begin{array}{r|rrrr} & -1&0&0&0\\\hline  -1&2&-1&-1&1 \\ 0&-1&-1&0&1\\
0&0&1&-1&0\\0&1&0&1&-1\\\end{array}\leq 1 \quad \quad I_{4422}^8
\equiv
\begin{array}{r|rrrr} & -2&-1&-1&0\\\hline  -2&1&1&2&1 \\ -1&1&2&-2&0\\
-1&2&-2&-1&1\\0&1&0&1&-2\\\end{array}\leq 0 \\\nonumber I_{4422}^9
\equiv
\begin{array}{r|rrrr} & -2&-1&-1&0\\\hline  -2&1&1&2&1 \\ -1&1&2&-2&0\\
-1&2&-2&-2&1\\0&1&0&1&-1\\\end{array}\leq 0 \quad \quad
I_{4422}^{10} \equiv
\begin{array}{r|rrrr} & -2&-1&-1&0\\\hline  -2&1&1&1&2 \\ -1&1&1&2&-2\\
-1&1&2&-2&-1\\0&2&-2&-1&-1\\\end{array}\leq 0 \\\nonumber
I_{4422}^{11} \equiv
\begin{array}{r|rrrr} & -2&-1&-1&0\\\hline  -2&1&1&1&2 \\ -1&1&0&2&-1\\
-1&1&2&-1&-1\\0&2&-1&-1&-1\\\end{array}\leq 0 \quad \quad
I_{4422}^{12} \equiv
\begin{array}{r|rrrr} & -2&-1&-1&0\\\hline  -2&1&1&1&2 \\ -1&1&-1&1&0\\
-1&1&1&2&-2\\0&2&0&-2&-1\\\end{array}\leq 0 \\\nonumber
I_{4422}^{13} \equiv
\begin{array}{r|rrrr} & -2&-1&-1&0\\\hline  -2&0&1&1&1 \\ -1&1&-2&1&1\\
-1&1&1&-1&1\\0&1&1&1&-1\\\end{array}\leq 0 \quad \quad
I_{4422}^{14} \equiv
\begin{array}{r|rrrr} & -2&-1&0&0\\\hline  -2&2&2&0&1 \\ -1&2&-1&1&-1\\
0&0&1&-1&-1\\0&1&-1&-1&0\\\end{array}\leq 0 \\\nonumber
I_{4422}^{15} \equiv
\begin{array}{r|rrrr} & -2&-1&0&0\\\hline  -2&2&1&1&1 \\ -1&1&-1&-1&1\\
0&1&-1&0&-1\\0&1&1&-1&-1\\\end{array}\leq 0 \quad \quad
I_{4422}^{16} \equiv
\begin{array}{r|rrrr} & -2&-1&0&0\\\hline  -2&2&0&1&1 \\ -1&0&1&-1&1\\
0&1&-1&-1&0\\0&1&1&0&-1\\\end{array}\leq 0 \\\nonumber
I_{4422}^{17} \equiv
\begin{array}{r|rrrr} & -2&-1&-1&-1\\\hline  -2&-1&1&2&2 \\ -1&1&-1&-1&2\\
-1&2&-1&2&-1\\-1&2&2&-1&0\\\end{array}\leq 0 \quad \quad
I_{4422}^{18} \equiv
\begin{array}{r|rrrr} & -2&-2&0&0\\\hline  -2&2&2&2&-1 \\ -2&2&1&-2&2\\
0&2&-2&-2&-2\\0&-1&2&-2&-1\\\end{array}\leq 0 \\\nonumber
I_{4422}^{19} \equiv
\begin{array}{r|rrrr} & -3&-2&0&0\\\hline  -3&2&2&1&2 \\ -2&2&-1&2&-2\\
0&1&2&-1&-1\\0&2&-2&-1&0\\\end{array}\leq 0 \quad \quad
I_{4422}^{20} \equiv
\begin{array}{r|rrrr} & -2&-2&-2&0\\\hline  -2&-1&1&1&2 \\ -2&1&-1&1&2\\
-2&1&1&-2&2\\0&2&2&2&-2\\\end{array}\leq 0 \\\ea

%\newpage

% ---------------------------------------------------------------------

\end{document}